\documentclass[conference,compsoc]{IEEEtran}
\ifCLASSOPTIONcompsoc
  \usepackage[nocompress]{cite}
\else
  \usepackage{cite}
\fi
\ifCLASSINFOpdf
\else
\fi
\usepackage{algorithmic}
\usepackage{ifdraft}

\usepackage[dvipdfmx]{color}
\usepackage[pdftex]{graphicx}

\usepackage{amsmath}
\usepackage{amsthm}
\usepackage{amssymb}
\usepackage{stmaryrd}
\usepackage{bm}
\usepackage{latexsym}
\usepackage{colonequals}

\usepackage{siunitx}

\usepackage{algorithm}
\usepackage{algorithmic}

\usepackage{etoolbox}
\makeatletter
\patchcmd{\@makecaption}
  {\scshape}
  {}
  {}
  {}
\makeatother

\begin{document}
%
\title{Graph Convolutional Network-based \\ Suspicious Communication Pair Estimation for Industrial Control Systems}

\author{
\IEEEauthorblockN{Tatsumi Oba}
\IEEEauthorblockA{
Panasonic Corp., Japan\\
oba.tatsumi@jp.panasonic.com}
\and
\IEEEauthorblockN{Tadahiro Taniguchi}
\IEEEauthorblockA{
Ritsumeikan Univ. \& Panasonic Corp., Japan\\
taniguchi@em.ci.ritsumei.ac.jp}}
\maketitle


%
\IEEEpeerreviewmaketitle


\begin{abstract}
Whitelisting is considered an effective security monitoring method for networks used in industrial control systems,
where the whitelists consist of observed tuples of the IP address of the server, 
the TCP/UDP port number, and IP address of the client (communication triplets). 
However, this method causes frequent false detections. 
To reduce false positives due to a simple whitelist-based judgment,
we propose a new framework for scoring communications to judge whether the 
communications not present in whitelists are normal or anomalous.
To solve this problem, we developed a graph convolutional network-based suspicious communication pair estimation using relational graph convolution networks,
and evaluated its performance. 
For this, we collected the network traffic of three factories owned by Panasonic Corporation, Japan.
The proposed method achieved a receiver operating characteristic area under the curve of 0.957,
which outperforms baseline approaches such as DistMult, a method that directly optimizes the node embeddings,
and heuristics, which score the triplets using first- and second-order proximities of multigraphs.
This method enables security operators to concentrate on significant alerts.

\end{abstract}

\section{Introduction}
\label{sec:introduction}
In industrial control systems (ICSs), significant effort has been made to generate
network communication whitelists to detect suspicious communications.
However, despite a large number of false alerts caused by a whitelist-based detection,
there have been no methods thus far that quantify the anomalies of communications not present on a whitelist.
Therefore, we propose a new framework for scoring communications not present on a whitelist 
to determine whether the communications are normal or anomalous.
We propose a graph convolutional network-based suspicious communication pair estimation (GCN SCOPE), a framework using relational graph convolutional networks (R-GCNs).
The proposed method regards the problem of scoring communications not present on a whitelist as a link prediction problem 
in multigraphs (graphs that are permitted to have multiple edges, that is, edges that have the same end nodes), 
where the nodes of graphs represent the IP addresses observed in the network,
and the edges of the graphs represent TCP/UDP port numbers used between the two IP addresses.

The importance of ICS protection, including their critical infrastructures such as power equipment 
and water processing facilities, has rapidly increased in recent years.
ICSs used to be considered safe against malware or cyberattacks because
they were isolated from enterprise IT systems or the Internet.
However, the growing requirements of remote monitoring, remote operations, and big data management 
have rapidly introduced the concept of the Internet of Things (IoT), and hence an increasing number of ICS networks are connecting to IT networks or the Internet.
As a result, many cases of malware infection in ICS networks, resulting in major damage, have been reported.
There have also been cases of social engineering attacks or attacks using removable devices.
Stuxnet\cite{Stuxnet}, which was discovered in 2010, is a type of malware that invades a stand-alone computer system 
through USB storage. Targeting Iran's nuclear facilities, it caused severe damage. 
Stuxnet is the first well-known example of an ICS being targeted. Moreover, because many general-purpose 
PCs are currently being used in ICSs, there are many cases in which they have been accidentally infected with malware 
that did not target the ICS system but forced a suspension of the operations.

Unlike a general IT environment, network-based monitoring solutions are preferred in an ICS.
Differing from most general IT environments, ICS environment requires a
continuous and stable device operation. 
A general PC receives frequent security updates, but in an ICS environment, down time during an updating procedure is not allowed. 
Therefore, in many cases, older system versions continued to run.
Moreover, because some devices are used for a long time (up to 10 to 20 years), 
security products may not be supported in certain cases.
 


In an ICS environment, although a whitelist-based detection method is thought to be effective,
it may cause a significant number of false detections.
With this method, whitelists consist of triplets (server IP address, TCP/UDP port number, and client IP address, hereinafter called a communication triplet), and alerts are raised when communication triplets that have not been previously observed
are noticed\cite{Barbosa13,stouffer11}. 
As a result of many false detections, security operators are forced to deal with false alerts, which makes the method impractical.

To solve this problem, we proposed a new framework, shown in Figure \ref{fig:overview}, to score the anomaly of 
unobserved communication triplets by learning communication triplets observed in the training data.
The output scores allow us to filter out unimportant detection events, and
focus only on fatal alerts.
Thus, the proposed method enables security operators to concentrate on significant alerts.

We propose GCN SCOPE, a method based on the framework,
utilizing R-GCNs\cite{R-GCN}.
The flow of the proposed method is shown in Figure \ref{fig:prediction_flow}.
An R-GCN is a model proposed by Schlichtkrull et al., that enables accurate link prediction in multigraphs.
The proposed method interprets the communication situation of ICS networks as multigraphs,
the nodes of which represent IP addresses, and the edges of which represent TCP/UDP port numbers, 
and estimates the possibility of the emergence of unobserved links.
This method enables us to avoid many false detections that cannot be avoided if we use
a simple whitelisting method, and correctly detect genuinely anomalous communications.

We independently collected network traffic for three manufacturing plants for 2 weeks each.
We use 1 week of data for training and 1 week of data for testing.
To investigate how well GCN SCOPE can distinguish between normal and anomalous triplets,
we use the test triplets as negative samples, and randomly extract triplets as positive samples,
and quantify the performance based on the distinguishability.

The proposed method achieved a receiver operating characteristic (ROC) area under the curve (AUC) of 0.957,
and outperforms baselines including DistMult and two other heuristics.

\noindent
\textbf{Contribution.}
The main contributions of this paper are as follows:
\begin{itemize}
\item We propose a new framework to quantify the anomalies of unobserved communication triplets
(consisting of tuples of the server IP address, TCP/UDP port number, and client IP address) by
learning communication triplets observed in the training data.
\item We developed a method based on the framework above using R-GCNs, and
demonstrated that this method can distinguish communication triplets observed as test data
from randomly extracted anomalous triplets while outperforming the accuracy of
baselines such as DistMult, which uses graph embedding such as in an R-GCN, and heuristics,
which score the triplets using the first- and second-order proximities of the graphs.
\end{itemize}

\noindent
\textbf{Paper Organization.}
The rest of this paper is organized as follows:
Related studies are described in Section \ref{sec:related_works}.
The problem statement is outlined in Section \ref{sec:problem_statement}.
Section \ref{sec:preliminaries} then describes the R-GCNs
used as components of our proposed approach.
We then show the details of our method in Section \ref{sec:proposed_method}.
Section \ref{sec:experiment} describes the data collected
and presents an evaluation of our experiment.
Finally, Section \ref{sec:conclusion} concludes this paper.



\section{Related Studies}
\label{sec:related_works}
Network-based monitoring methods are roughly divided into three types: 
signature-based detection, rule-based detection, and anomaly detection.
Although signature-based detection has few false positives, 
it can only detect known attacks. 
However, an increase in malware variants has decreased the signature detection rate.
Therefore, it is necessary to utilize other monitoring methods.
In rule-based detection, for example, information on the communication server and client pair, 
the protocols used, and the time of occurrence 
during a specific period are stored as a whitelist, 
and communications that are not present on the whitelist are detected.
Rules can also be manually created by referring to such information as specifications. 
Because this method stores only normal communications as a whitelist, 
it can detect unknown attacks. 
However, if the granularity of the rules to be created is too coarse, 
there is a risk of passing an attack, and if the granularity of the rule is too fine, 
there is a risk of causing many false positives.

Among the rule-based detection methods available, 
an approach using communication whitelists is considered to be 
particularly effective for ICS networks.
Barbosa et al. proposed a method that learns a set of tuples consisting of the server IP address, client IP address, TCP/UDP, 
and port number as a whitelist in a SCADA network,
and when the learning phase ends, an alert is generated for communications 
not present on the whitelist. \cite{Barbosa13}. 
However, this method causes many false detections in reality, 
and as a result, imposes a heavy load on security analysts, and is often impractical during an operation.
GCN SCOPE does not provide a binary output as in a simple whitelisting method
but quantifies an anomalous communication not present on a whitelist,
which allows us to focus only on genuine anomalous communications.
Choi et al. pointed out cases in which we cannot distinguish a client-server correctly,
and where the port numbers used by the clients vary and proposed a method to deal with such problems.
In this paper, we assume we can extract the server ports accurately and provide the same service.
In fact, we can extract such information from our dataset using Zeek (formerly Bro)\cite{Bro}.
Furthermore, the client port number tends to vary each time, and thus we ignore the client port number with our method.

The different types of anomaly detection methods include traffic anomaly detection methods
in which feature vectors are extracted from traffic of the entire network 
and anomalies are detected using machine learning. 
Yun et al. proposed an anomaly detection method in which the number of packets are observed from the traffic of the ICS network 
for each pair of source and destination devices for a certain period 
and are converted into feature vectors. An anomaly is detected when the pairs are far from the feature vectors observed during the training phase.
Kitsune\cite{Kitsune} efficiently extracted feature vectors 
related to traffic in real-time, such as information related only to the source IP, 
information related to the source and destination IPs, and information on the port number, and the
algorithm then detects anomalies using a reconstruction error of the autoencoder. 

However, the traffic anomaly detection methods described above work well
only when a large deviation (number of packets, data size, and transmission interval) 
occurs compared with usual traffic, and they cannot properly evaluate unobserved communication triplets.
In some cases, a cyberattack can be conducted with an extremely small number of packets or data sizes,
and in such cases, it cannot be detected by the traffic anomaly detection methods described above.

Other anomaly detection methods such as payload-based detection,
and methods based on the N-gram of the byte strings, are well-known \cite{PAYL,ANAGRAM,ZOE}.

However, these methods require a high calculation cost because it is necessary to analyze all packets in the network. 
Furthermore, such methods cannot detect attack packets with payloads similar to the byte string of usually observed payloads,
and do not perform well with a protocol containing many random byte sequences\cite{Hadziosmanovic12}. 
By contrast, the proposed method is extremely lightweight because it utilizes only the connectivity of the communication, 
and can detect the activity of an attacker without depending on the packet payload.





\section{Problem Statement}
\label{sec:problem_statement}
\begin{figure}[!tb]
  \begin{center}
    \includegraphics[width=\linewidth]{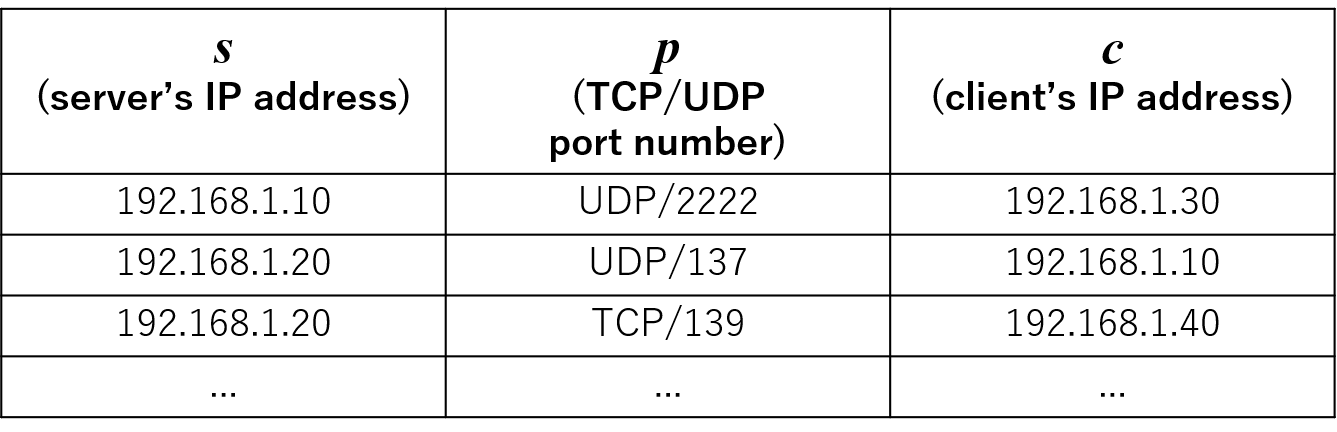}
  \caption{An example of a communication triplet}
  \label{fig:triplets}
  \end{center}
\end{figure}

In an ICS network, 
whitelisting is considered to be effective as a security monitoring method,
where the whitelists consist of tuples of the server IP address, 
TCP/UDP port number, and client IP address (communication triplets) that have appeared in the past,
and a rule-based detection method is applied that generates an alert when a communication triplet 
that has not been previously observed is discovered.
For example, communication triplets have the form shown in Figure \ref{fig:triplets}.
The reasons why this method is suitable for the ICS network are as follows:
\begin{itemize}
\item It operates on the network level, that is, there is no need to modify the devices.
\item In an ICS, new IP addresses and new communication pairs are unlikely to appear as compared with traditional IP networks,
where the number of legitimate connections is too large to be manageable.
\item It can handle proprietary protocols because it does not depend on the packet payload.
\item Although it is relatively easy to execute a lateral movement only within {\it observed IP pairs} in an ICS network 
because of the large number of valid communication IP pairs,
it is difficult to execute a lateral movement within {\it observed communication triplets}, which include TCP/UDP port numbers.
\end{itemize}

However, this method causes frequent false detections. 
The following patterns are factors that cause false positives:
\begin{enumerate}
\item Insufficient period of whitelist learning 
\item Fundamental changes in communication patterns
\item Unusual communications owing to non-steady operation (maintenance, troubleshooting, etc.)
\end{enumerate}
False positives from 1) may be avoided by sufficiently learning the whitelists for a lengthy period.
However, if we wait until the whitelist converges, we cannot detect anomalies for a long time.
Moreover, 2) may occur and the whitelist will need to be retrained before it converges.
Avoiding problem 3) is difficult in principle for whitelisting.

To reduce false positives caused by a simple whitelist-based judgment,
we distinguish whether non-whitelisted communications are caused as the result of a normal ICS environment 
operation or the result of an anomalous communication such as a cyberattack. 

\begin{figure*}[!tb]
  \begin{center}
    \includegraphics[width=\linewidth]{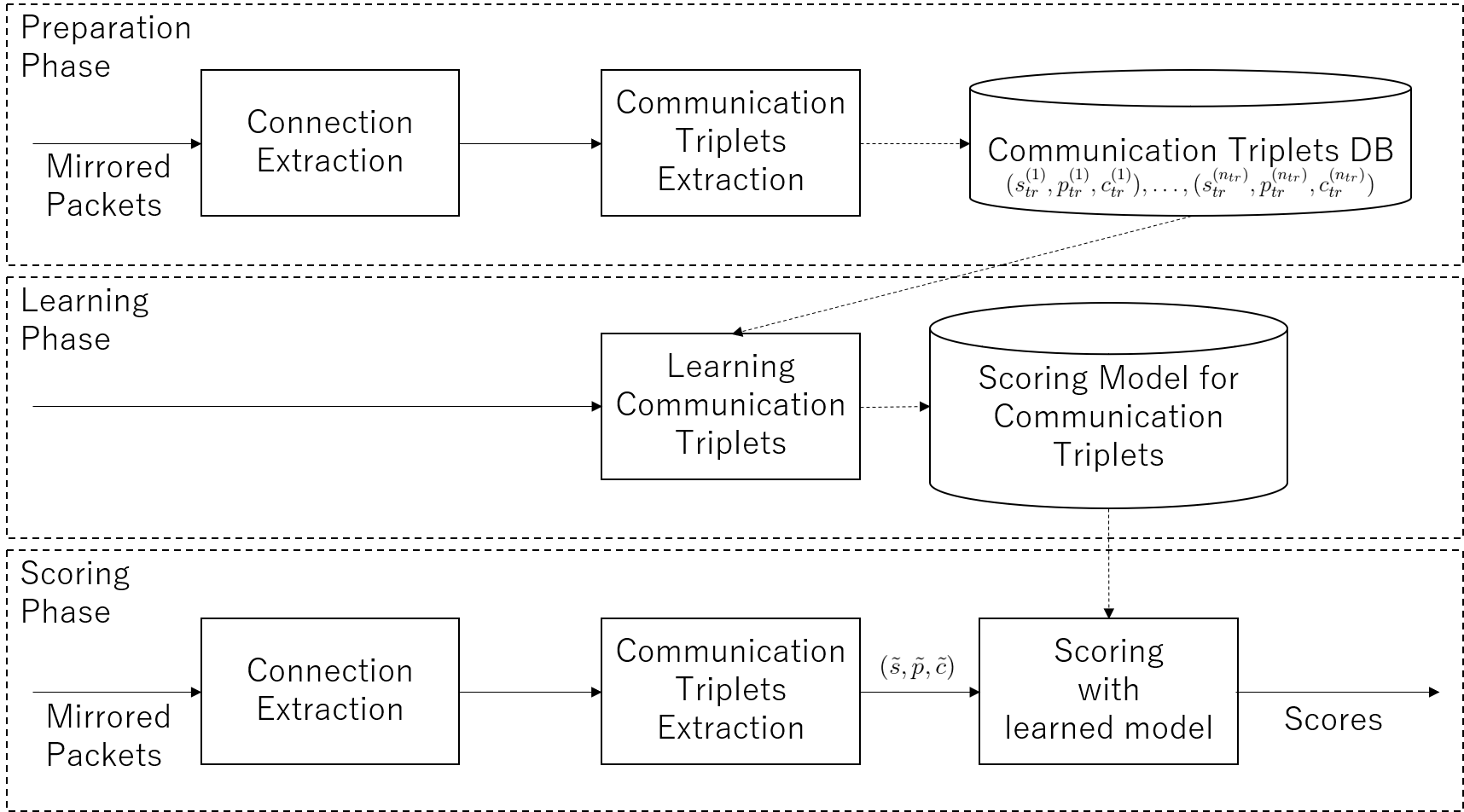}
  \caption{Framework for communication triplet scoring in this study.
  Communication triplet indicating tuples comprising the sets of $(s, p, c)$, where $s$ is the server's IP address, 
  $p$ is the TCP/UDP port number, and $c$ is the client IP address.}
  \label{fig:overview}
  \end{center}
\end{figure*}

To realize this, we propose a framework
consisting of the preparation phase, learning phase, and scoring phase, as in Figure \ref{fig:overview}.
Normal and anomalous communications can be separated
using an output score and the given threshold value, which is the case with whitelisting.
Therefore, the final output of the framework is the scores of the communication triplets.

We can quantify the anomalous communication triplets from normal whitelisted triplets
because none of the communications in an ICS network appear chaotic.
The protocols of communication may depend on the role of the device,
and some communications may be triggered by other communications.



\section{Preliminaries}
\label{sec:preliminaries}
Our method utilizes R-GCNs \cite{R-GCN} as 
the building components and we, therefore, describe such networks in this section.
R-GCNs are an extension of graph convolutional networks (GCNs) \cite{GCN} for multigraphs.
GCNs can be applied to semi-supervised node classification or a graph auto-encoder (GAE) \cite{VGAE}, and
R-GCNs can also be used for such applications.
Herein, we show the graph convolution calculation of R-GCNs,
an embedding learning method using R-GCNs given the learned triplets, 
and a link prediction method using a learned node and relation embeddings.

\subsection{Relational Graph Convolutional Networks}
R-GCNs are related to a class of neural networks operating on graphs,
and were developed specifically to deal with the highly multi-relational
data characteristics of realistic knowledge bases.

The following notations are introduced:
Directed and labeled multigraphs are denoted as $G=({\cal V}, {\cal E}, {\cal R})$
with nodes $v_i \in {\cal V}$ and labeled edges (relations) $(v_i, p, v_j) \in {\cal E}$,
where $p \in {\cal R}$ is a relation type.

Motivated by the architectures used by GCNs and other methods,
R-GCNs are defined through the following simple propagation model
for calculating the forward-pass update of the entity or node
denoted by $v_i$ in a relational multi-graph:
\begin{eqnarray}
\label{eq:rgcn}
h^{(l+1)}_i = \sigma\left(\sum_{p\in {\cal R}}\sum_{j \in {\cal N}_{i}^{p}}
\frac{1}{C_{i,p}}W_p^{(l)}h_j^{(l)} + W_0^{(l)}h_i^{(l)}\right),
\end{eqnarray}
where $h_i^{(l)} \in \mathbb{R}^{d^{(l)}}$ is the hidden state of node $v_i$
in the $l$-th layer of the neural network, ${\cal N}_i^p$ denotes
the set of neighbor indices of node $i$ under relation $p \in {\cal R}$,
and $W_p^{(l)}$ denotes the weight matrix for a simple linear transformation
depending on the relation $p$. In addition, $W_0$ is a weight matrix for a self-loop, and
$C_{i, p}$ is a problem-specific normalization constant that can either be
learned or chosen in advance (such as $C_{i,p} = |{\cal N}_i^p|$).

\subsection{Link Prediction Using R-GCN}
Link prediction deals with the prediction of new triplets (i.e., $(subject, relation, object)$).
Formally, a knowledge base is represented by a directed, labeled graph
$G=({\cal V}, {\cal E}, {\cal R})$. Rather than a full set of edges {\cal E},
only an incomplete subset $\hat{\cal E}$ is given.
The task is to assign scores $f(s, p, c)$ to possible edges $(s, p, c)$ to determine
how likely those edges are to belong to ${\cal E}$.
To tackle this problem, a graph auto-encoder model was introduced,
which is comprised of an entity encoder and a scoring function (decoder).
The encoder maps each entity $v_i \in {\cal V}$ to a real-valued vector $e_i \in \mathbb{R}^d$.
The decoder reconstructs edges of the graph relying on the vertex representations;
in other words, it scores $(s, p, c)$-triplets through a function
$f:\mathbb{R}^d \times {\cal R} \times \mathbb{R}^d \to \mathbb{R}$.
The representations through an R-GCN encoder can be computed as 
$e_i = h_i^{(L)}$, similar to the graph auto-encoder model introduced by Kipf and Welling\cite{VGAE}.

As the scoring function, DistMult factorization\cite{DistMult} is used,
which is known to perform well on standard link prediction benchmarks when
used on its own. In DistMult, every relation $p$ is associated with a {\it diagonal
matrix} $R_p \in \mathbb{R}^{d \times d}$, and a triple $(s,p,c)$ is scored as follows:
\begin{eqnarray}
f(s,p,c) = e_{s}^TR_pe_c.
\end{eqnarray}
An R-GCN decoder is based on DistMult, and does not explicitly model the asymmetry in the relation; hence,
$f(s,p,c) = e_{s}^TR_pe_c = e_{c}^TR_pe_s = f(c, p, s)$ is true.

The model is trained with negative sampling:
For each observed example $\omega$, negative examples are sampled.
The negative samples are sampled by randomly corrupting either the subject or object 
of each positive example.
The model is optimized for a cross-entropy loss to score observable triplets higher than
the negative triplets:
\begin{eqnarray}
\label{eq:learning_gae}
{\cal L} = - \frac{1}{(1+\omega)|\hat{\cal E}|} \sum_{(s,p,c,y)\in{\cal T}} y\log l\left( f(s,p,c)\right) + \nonumber \\
(1-y)\log (1-l\left(f(s,p,c)\right), 
\end{eqnarray}
where ${\cal T}$ is the total set of real and corrupted triplets, $l$ is the logistic sigmoid function,
and $y$ is an indicator set to $y=1$ for positive triplets and $y=0$ for negative triplets.


\section{Proposed Method}
\label{sec:proposed_method}
\begin{figure*}[!tb]
  \begin{center}
    \includegraphics[width=\linewidth]{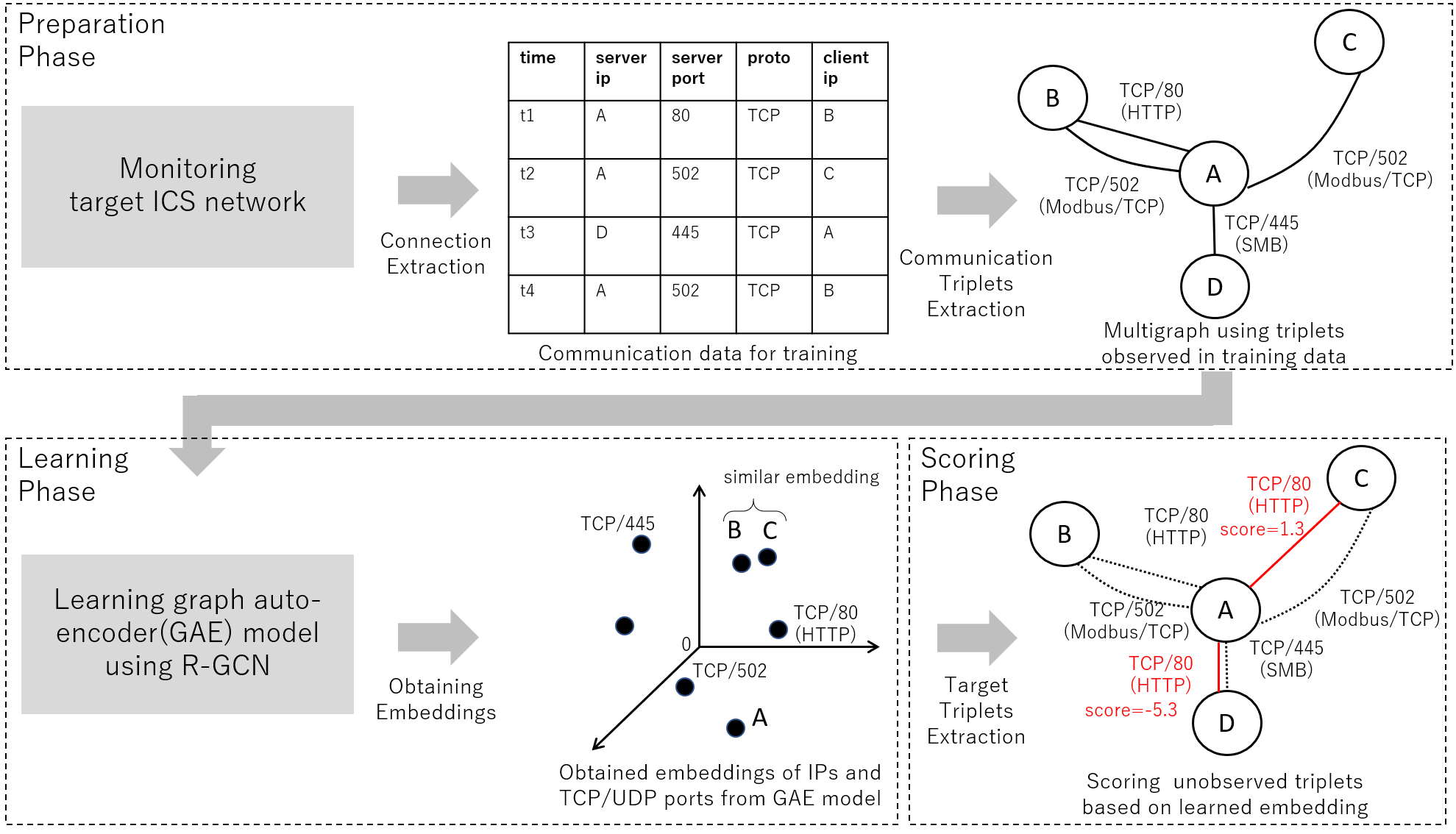}
  \caption{Link prediction flow of GCN SCOPE}
  \label{fig:prediction_flow}
  \end{center}
\end{figure*}

In this section, we show the key idea of the proposed method, GCN SCOPE,
and then describe the details of our approach.
The overall flow of the scoring of communication triplets with the proposed method
is shown in Figure \ref{fig:prediction_flow}.

\subsection{Key Idea}
As described in Section \ref{sec:problem_statement},
this study aims to detect anomalous communication triplets
based on normal communication triplets observed inside an ICS network.
This problem can be reduced to the problem of link prediction of multigraphs,
which is a task used to predict triplets
that are not clearly given but potentially exist with a high possibility.

The link between particular two devices can be estimated based on the {\it roles} of the devices as represented by their IP addresses,
where the {\it roles} indicate the types of device (such as HMI, PLC, RTU, Data historian, or SIS) and
the communication contents in the network.
We hypothesize that the {\it roles} can be recursively estimated by the connectivity among the {\it roles} of the neighbor devices.
We, therefore, focus on R-GCNs because they allow us to recursively extract embeddings expressing the {\it roles} of the devices.

For a {\it role} estimation, it is important to propagate the {\it role} information of the connected devices.
If a model does not take the features of each node into account and only considers the connectivity information,
the nodes cannot obtain the {\it role} information of the connected devices.
If a model tries to directly optimize the embeddings of each device such as in DistMult\cite{DistMult}
and not propagate the embeddings to the connected devices, then the embeddings of each node tend to overfit the training connections 
because the embeddings of each node are independently optimized for the connecting devices.
In fact, as we can see in Section \ref{sec:experiment},
the optimal embedding size of DistMult is smaller than
that of the proposed method.

\subsection{Scoring Communication Triplets}
We describe the details of our method based on 
the framework shown in Figure \ref{fig:overview} herein.

\subsubsection{Preparation Phase}
GCN SCOPE uses the information of the communication triplets observed in the target ICS networks.
The communication triplet is the tuple $(s, p, c)$
where $s$ indicates the server IP address, $p$ is a TCP/UDP port number, and $c$ represents a client IP address.
We have to extract all communication triplets $(s, p, c)$s observed during the learning period.
Let the set of all IP addresses, which emerge as the server IP addresses or client IP addresses, be ${\cal V}$,
the set of all the emerging TCP/UDP port numbers be ${\cal R}$, and
the set of all emerging communication triplets $(s, p, c)$ be $\hat{\cal E}$.
Note that the proposed method is based on an R-GCN, which assumes undirected multigraphs.
Therefore, if we observe one communication triplet $(s, p, c)$,
we consider the opposite triplet $(c, p, s)$ as having been observed as well.

The objective of our method is to score communication triplets in 
monitored ICS networks, and hence devices outside the target networks should be filtered out.
For example, if a target network allows some devices to communicate with the Internet through a gateway,
various IP addresses on the Internet appear that should be filtered out.

\subsubsection{Learning Phase}
In this phase, a graph autoencoder model using an R-GCN learns the communication triplets observed during the training period.
A part of the inputs of the learning phase is $({\cal V}, \hat{\cal E}, {\cal R})$ obtained during the preparation phase.
In addition to the inputs above, we have to provide some hyper parameters to a graph autoencoder model
such as the dropout rate, number of hidden layer units, L2 regularization weight,
the negative sampling rate, and regularization method (i.e., basis decomposition or block-diagonal decomposition;
for more details, see \cite{R-GCN}) and the parameters described therein.

We then run the learning algorithm of the graph autoencoder using the training triplets $\hat{\cal E}$
by optimizing the loss in Equation (\ref{eq:learning_gae}).
Finally, the model learns the parameters $R_{p_i} \in \mathbb{R}^{d \times d} \ (i \in \{1, \ldots, |{\cal R}|\})$ is the set of diagonal matrices of TCP/UDP port number embeddings 
and the weights of the R-GCN parameter 
$W_{p_i}^{(l)} \ (i \in \{1, \ldots, |{\cal R}|\}), l$ is the number of hidden layers$)$.
We can also obtain the embedding of the IP addresses,
$e_{i} \in \mathbb{R}^{d} \ (i \in \{1, \ldots, |{\cal V}|\})$, 
by calculating a forward-pass update in Equation (\ref{eq:rgcn}).
The overall learning algorithm is shown in Algorithm \ref{alg:learning}.

\begin{algorithm}
 \label{alg:learning}
 \caption{Communication triplet learning}
 \begin{algorithmic}[1] \label{alg:learning}
 \renewcommand{\algorithmicrequire}{\textbf{Input:}}
 \renewcommand{\algorithmicensure}{\textbf{Output:}}
 \REQUIRE ${\cal V}$: the set of observed IP addresses \\
          $\hat{\cal E}$: the set of trained triplets $(s, p, c)$ \\
          ${\cal R}$: the set of observed TCP/UDP port numbers \\
 \ENSURE  $e_{i} \in \mathbb{R}^{d} \ (i \in \{1, \ldots, |{\cal V}|\})$: the embeddings of IP addresses, \\
          $R_{p_j} \in \mathbb{R}^{d \times d} \ (j \in \{1, \ldots, |{\cal R}|\})$: the embeddings of TCP/UDP port numbers
 \STATE Learn $R_{p_i} \ (i\in \{1,\ldots,|{\cal R}|\})$ and $W_{p_j}^{(l)} \ (j\in \{1,\ldots,|{\cal R}|\}, l$ is the number of hidden layers$)$
        by minimizing the loss in Equation (\ref{eq:learning_gae})
 \STATE Calculate $e_{i} \in \mathbb{R}^{d} \ (i \in \{1, \ldots, |{\cal V}|\})$ by calculating the forward-pass update in Equation (\ref{eq:rgcn})
 \RETURN $e_{i} \ (i \in \{1, \ldots, |{\cal V}|\})$, $R_{p_j} \ (j \in \{1, \ldots, |{\cal R}|\})$
 \end{algorithmic} 
\end{algorithm}

\subsubsection{Scoring Phase}
In this phase, we consider the scoring target communication triplet $(\tilde{s}, \tilde{p}, \tilde{c})$.
We first check if $(\tilde{s}, \tilde{p}, \tilde{c})$ is included in the communication triplets 
$\hat{\cal E}$ observed during the learning period.
If $(\tilde{s}, \tilde{p}, \tilde{c})$ is included in $\hat{\cal E}$,
$(\tilde{s}, \tilde{p}, \tilde{c})$ is processed as normal and excluded from the scoring target.
Furthermore, if $\tilde{s}$ or $\tilde{c}$ is a new IP address or $\tilde{p}$ is a new TCP/UDP port number,
it is immediately processed as an anomalous triplet.
In addition, $(\tilde{s}, \tilde{p}, \tilde{c})$ is scored using an R-GCN only in cases other than the above two,
that is, cases $\tilde{s}$ and $\tilde{c}$ are in ${\cal V}$ and $\tilde{p}$ is in ${\cal R}$,
although the triplet $(\tilde{s}, \tilde{p}, \tilde{c})$ is not in $\hat{\cal E}$.
In this case, $(\tilde{s}, \tilde{p}, \tilde{c})$ are scored
using the embeddings $e_{i} \in \mathbb{R}^{d} \ (i \in \{1, \ldots, |{\cal V}|\})$ 
and $R_{p_i} \in \mathbb{R}^{d \times d} \ (i \in \{1, \ldots, |{\cal R}|\})$
obtained during the learning phase.
The embeddings of $(\tilde{s}, \tilde{p}, \tilde{c})$ are $(e_{\tilde s}, R_{\tilde p}, e_{\tilde c})$,
and the score of this triplet is calculated as $e_{\tilde s}^T R_{\tilde p} e_{\tilde c}$.
The overall scoring algorithm is shown in Algorithm \ref{alg:scoring}.

\begin{algorithm}
 \label{alg:scoring}
 \caption{Communication triplet scoring}
 \begin{algorithmic}[1] \label{alg:scoring}
 \renewcommand{\algorithmicrequire}{\textbf{Input:}}
 \renewcommand{\algorithmicensure}{\textbf{Output:}}
 \REQUIRE $(\tilde{s}, \tilde{p}, \tilde{c})$: scoring target triplet, \\
          ${\cal V}$: the set of observed IP addresses, \\
          $\hat{\cal E}$: the set of trained triplets $(s, p, c)$, \\
          ${\cal R}$: the set of observed TCP/UDP port numbers, \\
          $e_{p_i} \in \mathbb{R}^{d} \ (i \in \{1, \ldots, |{\cal V}|\})$: the set of node embeddings, \\
          $R_{p_i} \in \mathbb{R}^{d \times d} \ (i \in \{1, \ldots, |{\cal R}|\})$: the set of relation embeddings in a diagonal matrix
 \ENSURE  $score$: the score of $(\tilde{s}, \tilde{p}, \tilde{c})$
  \IF {$(\tilde{s}, \tilde{p}, \tilde{c}) \in \hat{\cal E}$}
  \STATE $score = \infty$
  \ELSIF {$\tilde{s} \not\in {\cal V}$ or $\tilde{c} \not\in {\cal V}$ or $\tilde{p} \not\in {\cal R}$}
  \STATE $score = - \infty$
  \ELSE
  \STATE $score = e_{\tilde s}^T R_{\tilde p} e_{\tilde c}$
  \ENDIF
 \RETURN $score$
 \end{algorithmic} 
\end{algorithm}



\section{Experiment}
\label{sec:experiment}
In this section, we first describe the dataset used in our experiments,
and then explain the baselines used as a comparison with our method.
Finally, we show the results of two experiments.

\subsection{Dataset}
In this paper, we used the network traffic of three factories owned by the company Panasonic for evaluation. 
The network monitoring of an ICS is generally conducted by collecting packets 
using a mirror port of an L2 switch,
and the datasets used in this paper are collected in the same manner.
Each factory produces different items, and the installed facilities, communication protocols, 
and network configurations are completely different.
Along with industrial protocols such as Modbus and Ethernet/IP, 
IT-based protocols such as NetBIOS, DNS, HTTP, HTTPS, FTP, SMB, RDP, SSH, MSSQL can be 
observed in these factories.

In these datasets, only unicast communications are the targets of learning and scoring,
and multicast and broadcast communications are excluded.
As the reason for excluding these communications, the incorporation of multicasting or broadcasting will result in the establishment of links with IP addresses
not specifically intended for communications.

The nature of each dataset is shown in Table \ref{tb:datasets}.
\begin{table}[tb]
\begin{tabular}{|l|l|l|l|l|} 
\hline
& Factory A& Factory B& Factory C \\ \hline
\# of IP addresses & 364 & 150 & 4109   \\ \hline
\# of TCP/UDP ports& 319 & 26 & 328 \\ \hline
\# of training triplets& 2241 & 2081 & 23993  \\ \hline
\# of test triplets & 764 & 558 & 4302 \\ \hline
\end{tabular}
\caption{Number of IP addresses, TCP/UDP port numbers, training triplets, and test triplets
in an ICS dataset.
The training and test triplets are only composed of unicast communications and are each collected during one week.}
\label{tb:datasets}
\end{table}
In Table \ref{tb:datasets}, the numbers of IP addresses, TCP/UDP port numbers, and training triplets
are obtained by counting the numbers of those that appeared during a specific week,
and test triplets are obtained as follows:
\begin{itemize}
\item Test triplets are composed only of triplets from data immediately following the week of training.
\item Triplets included in the training triplets are excluded from the test triplets.
\item Triplets with unobserved IP addresses or TCP/UDP port numbers are also excluded from the test triplets.
\end{itemize}


\subsection{Baselines}
We compare our method with three methods and uniform random scores.
The three methods are {\it DistMult}\cite{DistMult}, {\it first-order proximity-priority heuristic}
and {\it second-order proximity-priority heuristic}.

DistMult is a method for converting each node into the embeddings,
and the DistMult model is optimized using Equation \ref{eq:learning_gae}.
In other words, DistMult is equivalent to an R-GCN without graph convolution layers.

The two heuristics are methods considering first- and second-order proximities.
Many graph-embedding methods are designed to preserve this nature\cite{Goyal18}.
A first-order proximity to node $v_j$ from a view of node $v_i$ is higher
if more edges from $v_i$ to $v_j$ exist.
Let $s_i = [s_{i1}, \ldots, s_{in}]$ denote the first-order proximity
between $v_i$ and the other nodes. Then, the second-order proximity to node $v_j$ from a view of node $v_i$
is determined based on the similarity of $s_i$ and $s_j$.
In this paper, we use two variations as heuristics.
The first is a first-order proximity-first heuristic, and the other is a second-order proximity-first heuristic.
The former outputs a higher score to the node having higher first-order proximity,
and if two nodes have the same first-order proximity, it outputs a higher score to the node
having a higher second-order proximity.
The latter is the opposite method, which prioritizes second-order proximity.

\subsection{Results}
To evaluate GCN SCOPE, we conducted two types of experiments.
The first is a prediction performance experiment in which each method predicts 
the existence of communication triplets in the test dataset.
The other is a performance experiment in which each method
distinguishes whether each given communication triplet is in the test dataset 
or if anomalous triplets are randomly extracted.

With the proposed method, we utilize R-GCNs with two relational graph convolution layers,
and block diagonal decomposition regularization with a size of 10.
As the reason for using only two graph convolution layers, a
a large number of stacks of graph convolution layers is known to
degrade the performance\cite{QimaiLi18}.

The proposed method and DistMult both have multiple hyper parameters excluding those above.
Therefore, we split the data of factory A into training and validation data,
and search the hyper parameters using Bayes optimization with 50 iterations,
which achieves the best {\it mean reciprocal rank} (MRR)\cite{Bordes13} for the validation data.
The final hyper parameters adopted in the proposed method are as follows:
a dropout rate of 0.2, 100 hidden layer units,
an L2 regularization weight of 0.0, a learning rate of 0.01, and a
negative sampling rate of 10.
The final hyper parameters adopted in DistMult are as follows:
50 hidden layer units,
an L2 regularization weight of 0.01, a learning rate of 0.02,
and a negative sampling rate of 10.

\subsubsection{Link Prediction for Test Triplets}
We evaluate the performance of our method and the baselines using the {\it mean reciprocal rank} (MRR) and {\it Hits at n}(H@n),
where the models of each method learn the training triplets of each dataset and output the scores of the test triplets.
\[
MRR = \frac{1}{|Q|}\sum_{i=1}^{|Q|}\frac{1}{{\rm rank}_i},
\]
where ${\displaystyle {\text{rank}}_{i}}$ refers to the rank position of the correct answer for the i-th query.
Here, H@n is the proportion of the correct entities that are ranked within the top n.
This is the standard metric for link prediction methods.

The results are shown in Table \ref{tb:result1}.
Each metric is evaluated in the filtered settings.
\begin{table*}[tb]
\centering
\begin{tabular}{llllllllllllllllllll}
\hline
                                                                 & \multicolumn{4}{l}{Factory A}        &  & \multicolumn{4}{l}{Factory B}        &  & \multicolumn{4}{l}{Factory C}        &\\ \cline{2-5} \cline{7-10} \cline{12-15} \cline{17-20} 
                                                                 & MRR     & \multicolumn{3}{l}{Hits@n} &  & MRR     & \multicolumn{3}{l}{Hits@n} &  & MRR     & \multicolumn{3}{l}{Hits@n} &\\ \cline{1-5} \cline{7-10} \cline{12-15} \cline{17-20} 
Model                                                            &         & 1       & 3      & 10      &  &         & 1       & 3      & 10      &  &         & 1       & 3      & 10      &\\ \hline
\begin{tabular}[c]{@{}l@{}}GCN SCOPE (proposed)\end{tabular}   & {\bf 0.240}   & {\bf 0.172}   & {\bf 0.238}  & {\bf 0.366}   &  & {\bf 0.210}   & {\bf 0.108}   & {\bf 0.222}  & 0.395   &  & {\bf 0.291}   & {\bf 0.167}   & {\bf 0.423}  & {\bf 0.504}   & \\
DistMult\cite{DistMult}                                          & 0.177   & 0.096   & 0.198  & 0.334   &  & 0.161   & 0.047   & 0.158  & {\bf 0.435}   &  & 0.149   & 0.079   & 0.174  & 0.298   &\\
1st-order proximity-first heuristic                              & 0.182   & 0.122   & 0.189  & 0.312   &  & 0.063   & 0.006   & 0.041  & 0.159   &  & 0.192   & 0.120   & 0.263  & 0.287   &\\
2nd-order proximity-first heuristic                              & 0.168   & 0.101   & 0.179  & 0.277   &  & 0.055   & 0.009   & 0.031  & 0.116   &  & 0.151   & 0.056   & 0.244  & 0.286   & \\
Random                                                           & 0.016   & 0.001   & 0.005  & 0.021   &  & 0.040   & 0.007   & 0.022  & 0.070   &  & 0.002   & 0.000   & 0.001  & 0.002   &\\ \hline
\end{tabular}
\caption{Link prediction results for triplets observed in test data on real ICS datasets.}
\label{tb:result1}
\end{table*}
As shown in Table \ref{tb:result1}, GCN SCOPE outperforms the baselines in almost every case,
which means that GCN SCOPE shows a high performance in link prediction for the communication triplets of ICS networks.

\subsubsection{Distinction between Normal and Anomalous Links}
To investigate how well GCN SCOPE can distinguish between the normal and anomalous triplets,
we quantify the distinguishability of each method
using the AUC of the ROC curve.
We use the test triplets as negative samples, and randomly extract triplets as positive samples.

Although it is preferable to generate anomalous triplets based on known cyberattacks or a malware strategy, 
it is difficult to know the probability distribution, and thus random triplets are used instead.
We generate random communication triplets by choosing two different IP addresses and TCP/UDP port numbers separately and uniformly at random 
from those observed in the training data,
A total of 500 of these anomalous communication triplets are extracted in each dataset,
and each is composed only of triplets not included in the training and test triplets.

The evaluations are executed in two ways:
One method is a score-based evaluation (Table \ref{tb:result2_score_based}), 
and the other is a rank-based evaluation (Table \ref{tb:result2_rank_based}).
A score-based evaluation considers how a threshold judgment is executed using the original output scores.
A rank-based evaluation is based on the harmonic mean of $score = \frac{1}{rank_s} + \frac{1}{rank_c}$,
where $rank_s$ is calculated by finding the ranking of $(s,p,c)$ of all scores of the filtered communication triplet $(s,p,*)$,
and $rank_c$ is calculated by finding the ranking of $(s,p,c)$ of all scores of the filtered communication triplet $(*,p,c)$.
This is an evaluation metric based on the same idea as that used in an MRR.

\begin{table}[tb]
\centering
\begin{tabular}{ccccc}
\hline
Model                                                            & Factory A & Factory B & Factory C \\ \hline
\begin{tabular}[c]{@{}c@{}}GCN SCOPE (proposed)\end{tabular}     & {\bf 0.962} &{\bf0.914} &{\bf 0.996}\\
DistMult\cite{DistMult}                                          & 0.262       & 0.668     &  0.488    \\
1st-order proximity-first heuristic                              & 0.853       & 0.735     &  0.771    \\
2nd-order proximity-first heuristic                              & 0.820       & 0.632     &  0.769    \\
Random                                                           & 0.512       & 0.521     &  0.519    \\ \hline
\end{tabular}
\caption{ROC AUC of {\it score-based} link distinction on real ICS datasets.}
\label{tb:result2_score_based}
\end{table}

\begin{table}[tb]
\centering
\begin{tabular}{ccccc}
\hline
Model                                                            & Factory A & Factory B & Factory C \\ \hline
\begin{tabular}[c]{@{}c@{}}GCN SCOPE (proposed)\end{tabular}     & {\bf 0.903}     & {\bf 0.764}     & {\bf 0.989}     \\
DistMult\cite{DistMult}                                          & 0.710     & 0.745     & 0.900     \\
1st-order proximity-first heuristic                              & 0.767     & 0.708     & 0.767     \\
2nd-order proximity-first heuristic                              & 0.768     & 0.680     & 0.766     \\
Random                                                           & 0.554     & 0.521     & 0.532     \\ \hline
\end{tabular}
\caption{ROC AUC of {\it rank-based} link distinction on real ICS datasets.}
\label{tb:result2_rank_based}
\end{table}

From these evaluations, GCN SCOPE with a score-based judgment enables us to
distinguish normal triplets from anomalous triplets with high accuracy.
The performance of a score-based evaluation of DistMult is
significantly inferior to that of a rank-based evaluation,
the reason for which is considered to be based on the case of DistMult, in which all embeddings of the devices are optimized independently,
which causes a different scale of the scores in terms of the device.
By contrast, in the case of an R-GCN,
the embeddings are jointly optimized unlike with DistMult because of the effect of the graph convolution.



\section{Conclusion}
\label{sec:conclusion}
In this study, we developed GCN SCOPE, a method that learns the embeddings of IP addresses using
the learning of R-GCNs from observed communication triplets
(consisting of tuples of server IP addresses, TDP/UDP port numbers, and client IP addresses),
and scores unobserved communication triplets.
To the best of our knowledge, the proposed method is the first to quantify anomalous communication triplets
that have not previously been observed.
With the proposed method, a multigraph is constructed using communication triplets
observed in an ICS network, R-GCNs models are learned using the extracted multigraph, and
unobserved communication triplets are scored using learned R-GCNs models.
The proposed method achieved an average AUC of ROC curve of 0.957,
which outperforms the AUCs of comparative methods such as DistMult, a method that directly optimizes the node embeddings,
and heuristics, which score triplets using the first- and second-order proximities of multigraphs.
This means that the operators of a security operation center using communication whitelisting
are unleashed from the processing of large numbers of unimportant alerts.

In future studies, the performance changes will be observed using scoring functions when 
considering the direction of the edges, such as in TransE\cite{Bordes13} or ConvE\cite{ConvE}.
It also seems promising to use not only the edge types such as the TCP/UDP port numbers but
also the communication data size or communication time interval to enhance the accuracy of anomalous link detection.

\bibliographystyle{IEEEtran}
\bibliography{main}

\end{document}